\documentclass{jpsj-suppl}
\usepackage{txfonts} 

\renewcommand{\theenumi}{(\arabic{enumi})} 
\newcommand{\be}{\begin{equation}}
\newcommand{\ee}{\end{equation}}
\newcommand{\bea}{\begin{eqnarray}}
\newcommand{\eea}{\end{eqnarray}}
\newcommand{\beaa}{\begin{eqnarray*}}
\newcommand{\eeaa}{\end{eqnarray*}}

\newcommand{\abs}[1]{\vert{#1}\vert}
\def\Vec#1{\mbox{\boldmath $#1$}}

\def\Lap{{\mathop{\Delta}\limits^{(3)}}}

\def\be{\begin{equation}}
\def\ee{\end{equation}}
\def\bea{\begin{eqnarray}}
\def\eea{\end{eqnarray}}

\title{Large-scale magnetic fields from inflation in teleparallel gravity}

\author{Kazuharu \textsc{Bamba}$^{1, 2}$, Chao-Qiang \textsc{Geng}$^{1, 2, 3, 4}$ and Ling-Wei \textsc{Luo}$^{2, 4}$}

\inst{$^{1}$Kobayashi-Maskawa Institute for the Origin of Particles and the
Universe,
Nagoya University, Nagoya 464-8602, Japan \\
$^{2}$ Department of Physics, National Tsing Hua University, Hsinchu, 
Taiwan 300 \\
$^{3}$ Physics Division, National Center for Theoretical Sciences, Hsinchu, Taiwan 300 \\
$^{4}$ College of Mathematics \& Physics, Chongqing University of Posts \& Telecommunications, Chongqing, 400065, China
}

\email{bamba@kmi.nagoya-u.ac.jp}

\abst{Generation of large-scale magnetic fields in inflationary cosmology 
is studied in teleparallelism, 
where instead of the scalar curvature in general relativity, 
the torsion scalar describes the gravity theory. 
In particular, we investigate a coupling of 
the electromagnetic field to the torsion scalar during inflation, 
which leads to the breaking of conformal invariance of 
the electromagnetic field. 
We demonstrate that for a power-law type coupling, 
the current magnetic field strength of $\sim 10^{-9}$~G 
on 1 Mpc scale can be generated, if the backreaction effects and 
strong coupling problem are not taken into consideration. 
}

\kword{Particle-theory and field-theory models of the early Universe, 
Electric and magnetic fields, Modified theories of gravity}

\begin{document}
\maketitle

\section{Introduction}

The phenomenon that the accelerated expansion of the current universe 
has been suggested by 
various recent cosmological observations including Type Ia Supernovae, 
baryon acoustic oscillations, large scale structure, 
cosmic microwave background (CMB) radiation (e.g.,~\cite{Ade:2013lta}), 
and weak lensing. 
Provided that the universe is homogeneous, 
two representative procedures to 
account for the late time acceleration of the universe: 
(i) To introduce dark energy 
(for a recent review, see, e.g.,~\cite{Bamba:2012cp}) and 
(ii) To modify gravity such as $F(R)$ gravity~\cite{Nojiri:2010wj, Nojiri:2006ri}. 
Among the latter approaches, 
``teleparallelism''~\cite{Hehl:1976kj, Hayashi:1979qx} has recently been 
explored. 
It is written with the torsion scalar $T$, which is constructed by 
the Weitzenb\"{o}ck connection, and as a result, 
the action is expressed by $T$, although general relativity is 
described by the scalar curvature $R$, which is 
constructed by the Levi-Civita connection. 
In $f(T)$ gravity, the current accelerated expansion of the universe 
as well as inflation can be realized~\cite{Bengochea:2008gz, Linder:2010py, Bamba:2010iw, Bamba:2010wb}, as in $F(R)$ gravity. 
While, in galaxies magnetic fields with $\sim 10^{-6}$~G on 
the coherence scale $1$--$10$~kpc are observed. 
Furthermore, 
in clusters of galaxies, there exist 
large-scale magnetic fields with 
$10^{-7}$ -- $10^{-6}$~G on $10$~kpc--$1$~Mpc scale. 
The origin of cosmic magnetic fields including such 
large-scale magnetic fields in clusters of galaxies 
has not been well resolved yet. 
(for reviews, see, e.g.,~\cite{Grasso:2000wj, Giovannini:2003yn-2004rj, Kandus:2010nw, Durrer:2013pga}). 
Electromagnetic quantum fluctuations at the inflationary stage~\cite{Turner:1987bw} are considered to be the most natural mechanism to 
generate the large-scale magnetic fields. 
The reason is that inflation can make the scale of quantum fluctuations of 
the electromagnetic field extended. 
Moreover, according to the Quantum Electrodynamics of the curved space-time, 
thanks to one-loop vacuum-polarization effects~\cite{Drummond:1979pp}, 
a non-minimal gravitational coupling between 
the scalar curvature to the electromagnetic field can emerge. 
This breaks the conformal invariance of it, and 
eventually its quantum fluctuations are produced. 
Thus, large-scale magnetic fields can be generated at the current time~\cite{Turner:1987bw, Bamba:2003av, Bamba:2004cu, Martin:2007ue, Bamba:2006ga, Bamba:2007sx}. 
In this paper, we review our results obtained in Ref.~\cite{Bamba:2012mi}. 
In teleparallelism, 
we examine the generation of large-scale magnetic fields in inflationary 
universe. 
We consider the case that a non-minimal coupling between 
the torsion scalar to the electromagnetic field. 
We show that for a power-law coupling, 
the magnetic fields whose current strength is $\sim 10^{-9}$G on 1Mpc scale 
can be generated. 
We use units of $k_\mathrm{B} = c = \hbar = 1$ and 
adopt the Heaviside-Lorentz units of electromagnetism. 
In addition, we denote the gravitational constant $8 \pi G$ by 
${\kappa}^2 \equiv 8\pi/{M_{\mathrm{Pl}}}^2$ 
with the Planck mass of $M_{\mathrm{Pl}} = G^{-1/2} = 1.2 \times 10^{19}$GeV. 

\section{Non-minimal coupling of the Maxwell theory to the torsion scalar}

To begin with, we explain formulations of teleparallelism. 
In teleparallelism, 
orthonormal tetrad components $e_A (x^{\mu})$ are used. 
Here, at each point $x^{\mu}$ of a manifold, 
an index $A (= 0, 1, 2, 3)$ is 
for the tangent space. 
The metric $g^{\mu\nu}$ is described as 
$g_{\mu\nu}=\eta_{A B} e^A_\mu e^B_\nu$ with 
$\mu$ and $\nu$ ($= 0, 1, 2, 3$) coordinate 
indices on the manifold, and therefore 
$e_A^\mu$ make its tangent vector. 
The torsion and contorsion tensors are defined by 
$T^\rho_{\verb| |\mu\nu} \equiv e^\rho_A 
\left( \partial_\mu e^A_\nu - \partial_\nu e^A_\mu \right)$ 
and 
$K^{\mu\nu}_{\verb|  |\rho} 
\equiv -\left(1/2\right) 
\left(T^{\mu\nu}_{\verb|  |\rho} - T^{\nu \mu}_{\verb|  |\rho} - 
T_\rho^{\verb| |\mu\nu}\right)$, respectively. 
The torsion scalar is constructed as 
$T \equiv S_\rho^{\verb| |\mu\nu} T^\rho_{\verb| |\mu\nu}$ 
with 
$S_\rho^{\verb| |\mu\nu} \equiv \left(1/2\right)
\left(K^{\mu\nu}_{\verb|  |\rho}+\delta^\mu_\rho \ 
T^{\alpha \nu}_{\verb|  |\alpha}-\delta^\nu_\rho \ 
T^{\alpha \mu}_{\verb|  |\alpha}\right)$. 
In teleparallelism, the Lagrangian density 
is represented by the torsion scalar $T$, 
whereas that in general relativity, namely, the Einstein-Hilbert action, 
is written with the scalar curvature $R$. 
Consequently, in teleparallelism 
the action is given by 
\begin{equation}
S_\mathrm{Tel} = 
\int d^4x \abs{e} \left[ \frac{T}{2{\kappa}^2}  
+{\mathcal{L}}_{\mathrm{M}} \right], 
\label{eq:2.1}
\end{equation}
with $\abs{e}= \det \left(e^A_\mu \right)=\sqrt{-g}$ 
and ${\mathcal{L}}_{\mathrm{M}}$ the matter Lagrangian. 
We assume the flat 
Friedmann-Lema\^{i}tre-Robertson-Walker (FLRW) background 
with the metric 
$ds^2 = dt^2 - a^2(t) d{\Vec{x}}^2 
= a^2 (\eta) \left( -d \eta^2 + d{\Vec{x}}^2 \right)$, 
where $a$ is the scale factor and $\eta$ is the conformal time. 
In this case, 
$g_{\mu \nu}= \mathrm{diag} (1, -a^2, -a^2, -a^2)$ and  
$e^A_\mu = (1,a,a,a)$. 
{}From these relations, we obtain 
$T=-6H^2$ with $H \equiv \dot{a}/a$ the Hubble parameter, 
where the dot means the time derivative of $\partial/\partial t$. 

Next, an action of a non-minimal $I(T)$-Maxwell theory is 
described by 
\begin{equation}
S = 
\int d^{4}x \abs{e} 
\left(-\frac{1}{4}I(T)\, F_{\mu\nu}F^{\mu\nu}\right)\,, 
\label{eq:3.1}
\end{equation}
with 
$I(T)$ a function of $T$ and 
$F_{\mu\nu} = {\partial}_{\mu}A_{\nu} - {\partial}_{\nu}A_{\mu}$ 
the electromagnetic field-strength tensor, 
where $A_{\mu}$ is the $U(1)$ gauge field. 
{}From this action, the equation of motion for 
the electromagnetic field is given by 
$
-1/\sqrt{-g} {\partial}_{\mu} 
\left[ \sqrt{-g} I(T) F^{\mu\nu} 
\right] = 0
$. 
In the FLRW background, by using the Coulomb gauge of 
$A_0(t,\Vec{x}) = 0$ and ${\partial}^jA_j(t,\Vec{x}) =0$, 
the equation of motion for the $U(1)$ gauge field reads 
$
\ddot{A_i}(t,\Vec{x}) 
+ \left[ H + \left(\dot{I}(T) / I(T) \right) 
\right] \dot{A_i}(t,\Vec{x}) 
- \left( 1/a^2 \right) \Lap\, A_i(t,\Vec{x}) = 0
$ with $\Lap$ the Laplacian in the three-dimensional space. 
In addition, we quantize the $U(1)$ gauge field. 
The equation for the Fourier mode $A(k,t)$ becomes 
$
\ddot{A}(k,t) + \left( H + \dot{I}/I \right) \dot{A}(k,t) + 
\left(k^2/a^2\right) A(k,t) = 0
$
and the normalization condition is expresses as 
$
A(k,t){\dot{A}}^{*}(k,t) - {\dot{A}}(k,t){A^{*}}(k,t)
= i/\left(I a\right)$. 
With the conformal time $\eta$, the above equation 
for $A(k,t)$ is described as 
$A^{\prime \prime}(k,\eta) + 
\left( I^{\prime}/I \right) A^{\prime}(k,\eta) 
+ k^2 {A}(k,\eta) = 0$. 
Here, the prime denotes the derivative with respect to $\eta$ 
of $\partial/\partial \eta$. 
By applying the WKB approximation to the subhorizon scales 
and using the long-wavelength approximation on superhorizon scales, 
and matching these solutions at the horizon crossing, 
we can find an approximate analytic solution for the 
above equation of $A(k,t)$~\cite{Bamba:2006ga}. 
For the de Sitter space-time, 
we have $a=1/(-H\eta)$ and $-k\eta=1$ 
at the horizon-crossing time with $H=k/a$. 
In terms of the subhorizon (superhorizon) scales, 
$k|\eta| \gg 1$ ($k|\eta| \ll 1$). 
It is also satisfied for the generic slow-roll inflation, 
namely, almost exponential inflation. 
In the short-wavelength limit $k/\left(aH\right) \gg 1$,  
it is considered that the vacuum approaches the Minkowski vacuum 
asymptotically, so that 
the WKB solution for the subhorizon modes can become
$A_{\mathrm{in}} (k,\eta) = 
\left(1/\sqrt{2k}\right) I^{-1/2} e^{-ik\eta}$. 
On the other hand, by using the long-wavelength expansion in $k^2$, 
we acquire the solution on the superhorizon scales 
$A_{\mathrm{out}} (k,\eta)$. 
Matching it with the WKB solution for the subhorizon modes 
at the time of the horizon crossing of $\eta = \eta_k \approx 1/k$, 
we find 
$A_{\mathrm{out}} (k,\eta) = A_1 (k)$~\cite{Bamba:2006ga}, 
i.e., an approximate solution at the lowest order. 
Provided that after inflation, 
the instantaneous reheating occurs at $eta = {\eta}_{\mathrm{R}}$. 
The proper magnetic field is written by 
${B_i}^{\mathrm{proper}}(t,\Vec{x}) 
= a^{-1}B_i(t,\Vec{x}) = a^{-2}{\epsilon}_{ijk}{\partial}_j A_k(t,\Vec{x})
$ with $B_i(t,\Vec{x})$ the comoving magnetic field and 
${\epsilon}_{ijk}$ the totally antisymmetric tensor (${\epsilon}_{123}=1$). 
Hence, the magnetic field spectrum is represented as 
$|{B}^{\mathrm{proper}}(k,\eta)|^2  
=2\left(k^2/a^4\right)|A(k,\eta)|^2
=2\left(k^2/a^4\right)|A_1(k)|^2$. 
Here, the factor 2 coming from the two polarization degrees of freedom 
has been taken into consideration. 
The energy density of the Fourier modes of the magnetic fields reads 
${\rho}_B(k,\eta) = (1/2) |{B}^{\mathrm{proper}}(k,\eta)|^2 I(\eta)$. 
Using the phase-space density of $4 \pi k^3/(2\pi)^3$, 
the energy density of the magnetic fields per 
unit logarithmic interval of $k$ becomes 
$\rho_B(k,\eta) \equiv \left( 1/2 \right) \left[ 
4\pi k^3/\left(2\pi\right) \right] 
|{B}^{\mathrm{proper}}(k,\eta)|^2 I(\eta) 
=\left[ k|A_1(k)|^2/\left(2\pi^2\right) \right] 
\left(k^4/a^4\right) I(\eta)$. 
Thus, the density parameter and its spectral index 
are expressed as~\cite{Bamba:2006ga} 
$\Omega_B(k,\eta) = 
\left( \rho_B(k,\eta_{\mathrm{R}}) 
/\rho_\gamma(\eta_{\mathrm{R}}) \right) 
\left( I(\eta) / I(\eta_{\mathrm{R}}) \right)
= 
\left[ k^4/\left(T_{\mathrm{R}}^4 a_{\mathrm{R}}^4 \right) \right] 
\left[ 15k|A_1(k)|^2 / 
\left(N_{\mathrm{eff}}\pi^4\right)\right] I(\eta)$ 
and $n_B \equiv 
\left( d\ln\Omega_B(k) / d\ln k \right) = 
4+d \ln k|A_1(k)|^2 / d \ln k$. 
Here, we have used the following quantities at 
the reheating stage of $\eta = \eta_{\mathrm{R}}$: 
$\rho_\gamma (\eta_{\mathrm{R}})=N_{\mathrm{eff}} \left(\pi^2 /30 \right) 
T_{\mathrm{R}}^4$~\cite{K-T} is the energy density of radiation, 
$T_{\mathrm{R}}$ is the reheating temperature, 
$a_{\mathrm{R}}$ is the scale factor, 
and $N_{\mathrm{eff}}$ is the effective massless degrees of freedom 
thermalized during reheating. 
To demonstrate an estimation of the magnetic field strength 
quantitatively, the specific form of~\cite{Bamba:2006ga} 
$I(\eta)=I_*\left(\eta/\eta_{*}\right)^{-\beta}$ 
with $\eta_*$ a fiducial time during inflation, 
$I_*$ the value of $I$ at $\eta=\eta_*$, 
and $\beta$ a positive constant. 
Since $\beta >0$, $I$ monotonically increases at the inflationary stage. 
In this case, we have 
$k|A_1|^2=\left[1/ \left(2I(\eta_k)\right) \right] 
\left|1-\left(\beta+2i\right)/
\left[ 2 \left(\beta+1\right) \right] \right|^2 
\equiv \mathcal{A}/\left(2I(\eta_k)\right)$ 
with $\mathcal{A}$ a constant of $\mathcal{O}(1)$. 
By using $I(\eta_k) \propto k^\beta$ and $I(\eta_0) = 1$, 
the density parameter of the magnetic fields 
at the current time $\eta_0$ reads 
$\Omega_B(k,\eta_0)
= \left[ k^4/\left(T_{\mathrm{R}}^4 a_{\mathrm{R}}^4 \right) \right] 
\left[ 15\mathcal{A}/\left( 2N_{\mathrm{eff}}\pi^4I_* \right) \right] 
\left(\eta_k/\eta_*\right)^\beta$ 
with $n_B=4-\beta$. 
For a very small value of $I_*$, $\beta \sim 4$, namely, 
the spectrum becomes nearly scale-invariant, 
so that the generated large-scale magnetic fields can be strong. 
Also, the density parameter of the magnetic fields at the present time 
is expressed as~\cite{Bamba:2006ga} 
$\Omega_B(k,\eta_0)
= \mathcal{A} \left(N_\mathrm{eff}/1080\right) 
\left(T_{\mathrm{R}}/\tilde{M}_\mathrm{Pl}\right)^4
\left(-k\eta_{\mathrm{R}}\right)^{4-\beta}
\left(1/I(\eta_{\mathrm{R}}) \right)$, 
where we have used 
$a_{\mathrm{R}}^2\eta_{\mathrm{R}}^2 \approx H_{\mathrm{R}}^{-2}$ 
with $H_{\mathrm{R}}$ the Hubble parameter at the reheating stage 
and the Friedmann equation during reheating 
$3H_{\mathrm{R}}^2/ = 
\rho_\gamma (\eta_{\mathrm{R}}) /\tilde{M}_\mathrm{Pl}^2$ 
with $\tilde{M}_\mathrm{Pl} = M_\mathrm{Pl}/\sqrt{8\pi} = 1/\kappa$. 
Moreover, 
the term $\left(-k\eta_{\mathrm{R}}\right)$ is represented 
as~\cite{Bamba:2008my}  
$-k\eta_{\mathrm{R}} 
= k/\left(a_{\mathrm{R}} H_{\mathrm{R}} \right) \simeq 
\left(1.88/h\right) 10^4 \left(L/[\mathrm{Mpc}]\right) 
\left(T_{\mathrm{R}}/T_0\right) \left(H_0/H_{\mathrm{R}}\right) 
= 5.1 \times 10^{-25} N_\mathrm{eff}^{-1/2} 
\left(\tilde{M}_\mathrm{Pl} / T_{\mathrm{R}} \right) 
\left(L/[\mathrm{Mpc}]\right)^{-1}$. 
Here, in deriving the third equality, we have used 
$H_0^{-1}=3.0 \times 10^{3}\,h^{-1}$\,Mpc 
and $T \propto a^{-1}$. 
Moreover, when we obtain the last equality, 
we have used the Friedmann equation 
$3H_{\mathrm{R}}^2/ = 
\rho_\gamma (\eta_{\mathrm{R}}) /\tilde{M}_\mathrm{Pl}^2$ 
with 
$\rho_\gamma (\eta_{\mathrm{R}})=N_{\mathrm{eff}} \left(\pi^2 /30 \right) 
T_{\mathrm{R}}^4$, 
$T_0 = 2.73$\,K and $H_0=2.47 h \times 10^{-29}$\,K~\cite{K-T}  
with $h = 0.7$~\cite{Freedman:2000cf, Komatsu:2010fb}. 
The magnetic fields at the present time are written by 
$|B (\eta_0)|^2 = 2 \rho_B (\eta_0) = 2 \Omega_B (\eta_0,k)\,\rho_\gamma 
(\eta_0)$ 
with $\rho_\gamma (\eta_0) \simeq 2 \times 10^{-51}\,\mathrm{GeV}^4$ 
and $1\,\mathrm{G}=1.95 \times 10^{-20}\,\mathrm{GeV}^2$. 
Hence, we find~\cite{Bamba:2008my}
\begin{equation} 
|B(\eta_0, L)| = 2.7 \left[  \frac{7.2}{(5.1)^4 \pi} \right]^{\beta/8} 
\times 10^{-56+51\beta/4} 
N_\mathrm{eff}^{\left(\beta-4\right)/8} 
\sqrt{ \mathcal{A} \frac{I(\eta_0)}{I(\eta_{\mathrm{R}})} } 
\left( \frac{H_{\mathrm{R}}}{M_\mathrm{Pl}} \right)^{\beta/4}
\left( \frac{L}{[\mathrm{Mpc}]} \right)^{\beta/2-2} \, \mathrm{G}\,.
\label{eq:4.5}
\end{equation}
At the instantaneous reheating stage, 
$H_{\mathrm{R}}$ 
as $T_{\mathrm{R}} = \left[ 90/ \left(8 \pi^3 N_\mathrm{eff} \right) 
\right]^{1/4} \sqrt{M_\mathrm{Pl} H_{\mathrm{R}}}$. 
Furthermore, from tensor perturbations, 
we have a constraint on $H_{\mathrm{R}}$. 
{}From the Wilkinson Microwave Anisotropy Probe (WMAP) five year data 
in the anisotropy of the CMB radiation~\cite{Komatsu:2008hk}, 
we acquire 
$H_\mathrm{R} < 6.0 \times 10^{14}$GeV~\cite{Rubakov:1982df, Abbott:1984fp}.  
For a power-law inflation $a =a_0 \left( t/t_0 \right)^p$, 
where $p \gg 1$, and $a_0$ and $t_0$ are constants, 
with $\eta = \int \left( 1/a \right) dt$, we acquire 
$t/t_0 = \left[ a_0 t_0 \left( p-1 \right) \left( -\eta \right)
\right]^{-1/\left( p-1 \right)}$. 
If the coupling is a power-law type as 
$I(T) = \left( T/T_0 \right)^{n}$ with $T_0$ a current value of $T$ and 
$n (\neq 0)$ a non-zero constant, 
by using $T= -6H^2$ and $H= p/t$, 
we have 
$I(\eta) = \left(-6/T_0 \right)^n \left( p/t_0 \right)^{2n} 
\left[ a_0 t_0 \left( p-1 \right) \right]^{2n/\left( p-1 \right)} 
\left( -\eta \right)^{2n/\left( p-1 \right)}$ 
and 
$\beta=-\left( 2 n \right)/\left(p-1\right)$. 
When $N_\mathrm{eff} = 100$, $H_{\mathrm{R}} = 1.0 \times 10^{14}$GeV, 
i.e., $T_{\mathrm{R}} = 8.6 \times 10^{15}$GeV, 
$L = 1$Mpc, $\mathcal{A} = 1$, $I(\eta_{\mathrm{R}}) = I(\eta_0)$, 
and $\beta = 4.2$ (namely, $p=10$ and $n=-18.9$), 
we obtain 
$|B(\eta_0, L = 1\,\mathrm{Mpc})| = 2.5 \times 10^{-9} \, \mathrm{G}$. 
Furthermore, for the above values except 
$H_{\mathrm{R}} = 1.0 \times 10^{10}$GeV, i.e., 
$T_{\mathrm{R}} = 8.6 \times 10^{13}$GeV, 
and $\beta = 4.6$ (that is, $p=10$ and $n=-19.7$), 
we find 
$|B(\eta_0, L = 1\,\mathrm{Mpc})| = 2.3 \times 10^{-9} \, \mathrm{G}$. 
We caution that 
$I(T)$ evolves at the inflationary stage, and after instantaneous reheating 
it does not change any more as $I(\eta_{\mathrm{R}}) = I(\eta_0)$.

\section{Summary}

We have explored the 
generation of large-scale magnetic fields from inflation 
in teleparallelism. 
We have shown that 
when a power-law type coupling between the torsion scalar and 
the electromagnetic field exists at the inflationary stage,  
the magnetic fields with those strength of $\sim 10^{-9}$~G 
on 1 Mpc scale can be generated. 
If there are the large-scale magnetic fields with such a strength, 
by the adiabatic compression process without galactic dynamo, 
the magnetic fields observed in galaxies and clusters of galaxies 
can be explained. 
Finally, we should caution that in this work, the backreaction effects~\cite{Demozzi:2009fu, Kanno:2009ei, Suyama:2012wh, Fujita:2012rb}, namely, 
the energy density of the so-called inflaton field, which is a scalar field responsible for the realization of inflation, must be much larger than that of 
the electromagnetic fields during inflation, 
and strong coupling problem are not included.


\end{document}